\documentclass[twocolumn]{aastex62}

\graphicspath{{./}{figures/}}
\DeclareUnicodeCharacter{0301}{*************************************}

\shorttitle{TOI-1420b}
\shortauthors{Yoshida et al.}
\usepackage{physunits}
\usepackage{graphicx}
\usepackage{lineno}
\usepackage[T1]{fontenc}
\usepackage{natbib}
\begin{document}

\title{TESS Spots a Super-Puff: The Remarkably Low Density of TOI-1420b}

\author[0000-0003-4015-9975]{Stephanie Yoshida}
\affiliation{Center for Astrophysics \textbar \ Harvard \& Smithsonian, 60 Garden Street, Cambridge, MA 02138, USA}

\author[0000-0003-2527-1475]{Shreyas Vissapragada}
\altaffiliation{51 Pegasi b Fellow}
\affiliation{Center for Astrophysics \textbar \ Harvard \& Smithsonian, 60 Garden Street, Cambridge, MA 02138, USA}

\author[0000-0001-9911-7388]{David W. Latham}
\affiliation{Center for Astrophysics \textbar \ Harvard \& Smithsonian, 60 Garden Street, Cambridge, MA 02138, USA}

\author[0000-0001-6637-5401]{Allyson Bieryla}
\affiliation{Center for Astrophysics \textbar \ Harvard \& Smithsonian, 60 Garden Street, Cambridge, MA 02138, USA}

\author[0000-0002-5113-8558]{Daniel P. Thorngren}
\altaffiliation{Davis Postdoctoral Fellow}
\affil{Department of Physics \& Astronomy, Johns Hopkins University, Baltimore, MD, USA}

\author[0000-0003-3773-5142]{Jason D. Eastman}
\affiliation{Center for Astrophysics \textbar \ Harvard \& Smithsonian, 60 Garden Street, Cambridge, MA 02138, USA}

\author[0000-0003-3204-8183]{Mercedes L\'{o}pez-Morales}
\affiliation{Center for Astrophysics \textbar \ Harvard \& Smithsonian, 60 Garden Street, Cambridge, MA 02138, USA}

%start of alphabetic section

\author[0000-0003-1464-9276]{Khalid Barkaoui}
\affiliation{Astrobiology Research Unit, Universit\'e de Li\`ege, 19C All\'ee du 6 Ao\^ut, 4000 Li\`ege, Belgium}
\affil{Department of Earth, Atmospheric, and Planetary Sciences, Massachusetts Institute of Technology, Cambridge, MA 02139, USA}
\affiliation{Instituto de Astrof\'isica de Canarias (IAC), Calle V\'ia L\'actea s/n, 38200, La Laguna, Tenerife, Spain}

\author[0000-0002-5627-5471]{Charles~Beichman}
\affil{Infrared Processing and Analysis Center, California Institute of Technology, 1200 East California Blvd, Pasadena, CA 91125, USA}

\author{Perry Berlind}
\affiliation{Center for Astrophysics \textbar \ Harvard \& Smithsonian, 60 Garden Street, Cambridge, MA 02138, USA}

\author[0000-0003-1605-5666]{Lars~A.~Buchave}
\affil{DTU Space, National Space Institute, Technical University of Denmark, Elektrovej 327, DK-2800 Lyngby, Denmark}

\author{Michael L. Calkins}
\affiliation{Center for Astrophysics \textbar \ Harvard \& Smithsonian, 60 Garden Street, Cambridge, MA 02138, USA}

\author[0000-0002-5741-3047]{David~R.~Ciardi}
\affil{Infrared Processing and Analysis Center, California Institute of Technology, 1200 East California Blvd, Pasadena, CA 91125, USA}

\author[0000-0001-6588-9574]{Karen A.\ Collins}
\affiliation{Center for Astrophysics \textbar \ Harvard \& Smithsonian, 60 Garden Street, Cambridge, MA 02138, USA}

\author[0000-0003-1784-1431]{Rosario~Cosentino}
\affiliation{Fundaci\'on Galileo Galilei-INAF, Rambla Jos\'e Ana Fernandez P\'erez 7, 38712 Bre\~na Baja, Tenerife, Spain}

\author{Ian~J.~M.~Crossfield}
\affil{Department of Physics and Astronomy, University of Kansas, Lawrence, KS, USA}

\author[0000-0002-8958-0683]{Fei Dai}
\altaffiliation{NASA Sagan Fellow}
\affil{Division of Geological and Planetary Sciences, California Institute of Technology, 1200 East California Blvd, Pasadena, CA 91125, USA}
\affiliation{Department of Astronomy, California Institute of Technology, Pasadena, CA 91125, USA}

\author[0000-0003-0741-7661]{Victoria DiTomasso}
\affiliation{Center for Astrophysics \textbar \ Harvard \& Smithsonian, 60 Garden Street, Cambridge, MA 02138, USA}

\author{Nicholas Dowling}
\affil{University Observatory, Ludwig Maximilian University, Scheinerstrasse 1, D-81679 Munich, Germany}

\author{Gilbert~A.~Esquerdo}
\affiliation{Center for Astrophysics \textbar \ Harvard \& Smithsonian, 60 Garden Street, Cambridge, MA 02138, USA}

\author[0000-0002-6482-2180]{Raquel For\'{e}s-Toribio}
\affil{Departamento de Astronom\'{\i}a y Astrof\'{\i}sica, Universidad de Valencia, E-46100 Burjassot, Valencia, Spain} 
\affil{Observatorio Astron\'omico, Universidad de Valencia, E-46980 Paterna, Valencia, Spain}

\author[0000-0003-4702-5152]{Adriano Ghedina}
\affiliation{Fundaci\'on Galileo Galilei-INAF, Rambla Jos\'e Ana Fernandez P\'erez 7, 38712 Bre\~na Baja, Tenerife, Spain}

\author[0000-0003-2228-7914]{Maria V. Goliguzova}
\affiliation{Sternberg Astronomical Institute Lomonosov Moscow State University 119992, Moscow, Russia, Universitetskii prospekt, 13}

\author{Eli Golub}
\affil{NSF’s National Optical-Infrared Astronomy Research Laboratory, 950 N. Cherry Ave., Tucson, AZ 85719, USA}

\author[0000-0002-9329-2190]{Erica~J.~Gonzales}
\altaffiliation{NSF Graduate Research Fellow}
\affil{University of California, Santa Cruz, 1156 High Street, Santa Cruz CA 95065, USA}

\author[0000-0001-9927-7269]{Ferran Grau Horta}
\affiliation{Observatori de Ca l'Ou, Carrer de dalt 18, Sant Martí Sesgueioles 08282, Barcelona, Spain}

\author[0000-0002-3985-8528]{Jesus Higuera}
\affil{NSF’s National Optical-Infrared Astronomy Research Laboratory, 950 N. Cherry Ave., Tucson, AZ 85719, USA}

\author[0000-0001-7227-2556]{Nora Hoch}
\affil{Department of Astronomy, Wellesley College, Wellesley, MA 02481, USA}

\author[0000-0003-1728-0304]{Keith Horne}
\affiliation{SUPA Physics and Astronomy, University of St~Andrews, Fife, KY16 9SS Scotland, UK}

\author{Steve~B.~Howell}
\affil{NASA Ames Research Center, Moffett Field, CA 94035 USA}

\author{Jon~M.~Jenkins}
\affil{NASA Ames Research Center, Moffett Field, CA 94035 USA}

\author[0000-0003-3906-9518]{Jessica Klusmeyer}
\affil{NSF’s National Optical-Infrared Astronomy Research Laboratory, 950 N. Cherry Ave., Tucson, AZ 85719, USA}

\author{Didier Laloum}
\affil{Soci\'et\'e Astronomique de France, 3 Rue Beethoven, 75016 Paris, France}

\author[0000-0001-6513-1659]{Jack J. Lissauer}
\affiliation{Space Science \& Astrobiology Division, MS 245-3, NASA Ames Research Center, Moffett Field, CA 94035, USA}

\author[0000-0002-9632-9382]{Sarah E. Logsdon}
\affil{NSF’s National Optical-Infrared Astronomy Research Laboratory, 950 N. Cherry Ave., Tucson, AZ 85719, USA}

\author[0000-0002-6492-2085]{Luca~Malavolta}
\affiliation{Dipartimento di Fisica e Astronomia Galileo Galilei, Universit ́a di Padova, Vicolo dell’Osservatorio 3, 35122 Padova, Italy}

\author{Rachel~A.~Matson}
\affil{The United States Naval Observatory, 3450 Massachusetts Avenue, NW, Washington, DC 20392, USA}

\author[0000-0003-0593-1560]{Elisabeth~C.~Matthews}
\affil{Max-Planck-Institut für Astronomie, Königstuhl 17, D-69117 Heidelberg, Germany}

\author[0000-0001-9504-1486]{Kim K. McLeod}
\affil{Department of Astronomy, Wellesley College, Wellesley, MA 02481, USA}

\author{Jennifer V.~Medina}
\affiliation{Space Telescope Science Institute, 3700 San Martin Drive, Baltimore, MD, 21218, USA}

\author{Jose A. Mu\~noz}
\affil{Departamento de Astronom\'{\i}a y Astrof\'{\i}sica, Universidad de Valencia, E-46100 Burjassot, Valencia, Spain}
\affil{Observatorio Astron\'omico, Universidad de Valencia, E-46980 Paterna, Valencia, Spain}  

\author[0000-0002-4047-4724]{Hugh~P.~Osborn}
\affiliation{Department of Physics and Kavli Institute for Astrophysics and Space Research, Massachusetts Institute of Technology, Cambridge, MA 02139, USA}
\affiliation{NCCR/Planet-S, Universität Bern, Gesellschaftsstrasse 6, 3012 Bern, Switzerland}

\author[0000-0003-1713-3208]{Boris~Safonov}
\affiliation{Sternberg Astronomical Institute Lomonosov Moscow State University 119992, Moscow, Russia, Universitetskii prospekt, 13}

\author[0000-0001-5347-7062]{Joshua~Schlieder}
\affil{NASA Goddard Space Flight Center, 8800 Greenbelt Road, Greenbelt, MD 20771, USA}

\author{Michael Schmidt}
\affil{University Observatory, Ludwig Maximilian University, Scheinerstrasse 1, D-81679 Munich, Germany}

\author[0000-0001-9580-4869]{Heidi Schweiker}
\affil{NSF’s National Optical-Infrared Astronomy Research Laboratory, 950 N. Cherry Ave., Tucson, AZ 85719, USA}

\author[0000-0002-6892-6948]{Sara Seager}
\affil{Department of Earth, Atmospheric, and Planetary Sciences, Massachusetts Institute of Technology, Cambridge, MA 02139, USA}
\affil{Department of Physics and Kavli Institute for Astrophysics and Space Research, Massachusetts Institute of Technology, Cambridge, MA 02139, USA}
\affil{Department of Aeronautics and Astronautics, Massachusetts Institute of Technology, Cambridge, MA 02139, USA}

\author[0000-0002-7504-365X]{Alessandro Sozzetti}
\affil{INAF – Osservatorio Astrofisico di Torino, Via Osservatorio, 20, 10025 Pino Torinese TO, Italy}

\author{Gregor Srdoc}
\affil{Kotizarovci Observatory, Sarsoni 90, 51216 Viskovo, Croatia}

\author[0000-0001-7409-5688]{Gu{\dj}mundur Stef{\'a}nsson}
\altaffiliation{NASA Hubble Fellow}
\affil{Department of Astrophysical Sciences, Princeton University, 4 Ivy Lane, Princeton, NJ 08540, USA}

\author[0000-0003-0647-6133]{Ivan A. Strakhov}
\affiliation{Sternberg Astronomical Institute Lomonosov Moscow State University 119992, Moscow, Russia, Universitetskii prospekt, 13}

\author[0009-0008-5145-0446]{Stephanie Striegel}
\affil{SETI Institute, Mountain View, CA 94043 USA}
\affil{NASA Ames Research Center, Moffett Field, CA 94035 USA}

\author{Joel Villase{\~ n}or}
\affiliation{Department of Physics and Kavli Institute for Astrophysics and Space Research, Massachusetts Institute of Technology, Cambridge, MA 02139, USA}

\author[0000-0002-4265-047X]{Joshua N.\ Winn}
\affil{Department of Astrophysical Sciences, Princeton University, 4 Ivy Lane, Princeton, NJ 08540, USA}

\begin{abstract}
    We present the discovery of TOI-1420b, an exceptionally low-density ($\rho = 0.08\pm0.02$~g~cm$^{-3}$) transiting planet in a $P = 6.96$~day orbit around a late G dwarf star. Using transit observations from TESS, LCOGT, OPM, Whitin, Wendelstein, OAUV, Ca l'Ou, and KeplerCam along with radial velocity observations from HARPS-N and NEID, we find that the planet has a radius of $R_\mathrm{p} = 11.9\pm0.3R_\Earth$ and a mass of $M_\mathrm{p} = 25.1\pm3.8M_\Earth$. TOI-1420b is the largest-known planet with a mass less than $50M_\Earth$, indicating that it contains a sizeable envelope of hydrogen and helium. We determine TOI-1420b's envelope mass fraction to be $f_\mathrm{env} = 82^{+7}_{-6}\%$, suggesting that runaway gas accretion occurred when its core was at most $4-5\times$ the mass of the Earth. TOI-1420b is similar to the planet WASP-107b in mass, radius, density, and orbital period, so a comparison of these two systems may help reveal the origins of close-in low-density planets. With an atmospheric scale height of 1950~km, a transmission spectroscopy metric of 580, and a predicted Rossiter-McLaughlin amplitude of about $17$~m~s$^{-1}$, TOI-1420b is an excellent target for future atmospheric and dynamical characterization.
\end{abstract}

%\keywords{exoplanets}

\section{Introduction}
Since the discovery of 51 Pegasi b nearly thirty years ago \citep{Mayor1995}, over 5000 exoplanets have been detected to date. Many of these planets challenge our intuition from the Solar system. For instance, the Kepler mission \citep{Borucki2010} revealed that sub-Neptunes and super-Earths (with $1R_\Earth<R_\mathrm{p}<4 R_\Earth$ and $P < 100$~days) occur around $30-60\%$ of Sun-like stars \citep[e.g.][]{Latham2011}, despite not having a direct counterpart within the Solar System. The Solar System also exhibits a clear distinction between the ice giants ($M_\mathrm{p} \lesssim 20M_\Earth$) and the gas giants ($M_\mathrm{p} \gtrsim 100M_\Earth$). Many planets have now been detected with masses between that of Neptune and Saturn, although they are less common than sub-Neptunes and more challenging to detect than gas giants \citep{Petigura2018}.

\begin{deluxetable*}{cccc}
\tablecaption{Summary of Stellar Parameters for TOI-1420 \label{tab:stellar}}
\tablecolumns{4}
\tablehead{
\colhead{Parameters} &
\colhead{Description (Units)} &
\colhead{Values} &
\colhead{Source}
}
\startdata
Main Identifiers: \\
TOI & ... & 1420 & TESS Mission \\
TIC & ... & 321857016 & TIC \\
Tycho-2 & ... & 4261-149-1 & Tycho-2 \\
2MASS & ... & J21314590+6620556 & 2MASS \\
AllWISE & ... & J213145.99+662056.2 & AllWISE \\
Gaia DR3 & ... & 2221164434736927360 & Gaia DR3 \\
\hline
Coordinates \& Proper Motion: \\
$\alpha_{J2000}$ & Right Ascension (RA) & 21:31:45.917 & Gaia DR3 \\ 
$\delta_{J2000}$ & Declination (Dec) & +66:20:55.925 & Gaia DR3 \\
$\mu_{\alpha}$ & RA Proper Motion (mas/yr) & 45.482 ± 0.013 & Gaia DR3 \\
$\mu_{\delta}$ & Dec Proper Motion (Dec, mas/yr) & 31.874 ± 0.012 & Gaia DR3 \\
$\varpi$ & Parallax (mas) & 4.9134 ± 0.0105 & Gaia DR3 \\
$d$ & Distance (pc) & 201.84 ± 0.43 & Gaia DR3 \\
% $P$ & Period (days) & 6.956 & TESS Input Catalog \\
% $T_c$ & Epoch (Julian days) & 2458745.30557 & TESS Input Catalog \\
%\textbf{Stellar Parameters:} \\
%$M_*$ & Mass ($M_\odot$) & 0.935 & TESS Input Catalog \\
%$R_*$ & Radius ($R_\odot$) & 0.940 & TESS Input Catalog \\
\hline
Magnitudes: \\
$G$	 & Gaia $G$ Magnitude & 11.7323 ± 0.0002 & Gaia DR3 \\
$B_P$ & Gaia $B_P$ Magnitude & 12.1338 ± 0.0007 & Gaia DR3 \\
$R_P$ & Gaia $R_P$ Magnitude & 11.1707 ± 0.0006	& Gaia DR3 \\
$T$ & TESS Magnitude & 11.229 ± 0.006 & TIC \\
$J$ & 2MASS $J$ Magnitude & 10.557 ± 0.022 & 2MASS \\
$H$ & 2MASS $H$ Magnitude & 10.191 ± 0.021 & 2MASS \\
$K_s$ & 2MASS $K_s$ Magnitude & 10.119 ± 0.022 & 2MASS \\
$W1$ & WISE $W1$ Magnitude & 10.059 ± 0.023 & AllWISE \\
$W2$ & WISE $W2$ Magnitude & 10.120 ± 0.021 & AllWISE \\
$W3$ & WISE $W3$ Magnitude & 10.084 ± 0.044 & AllWISE \\
\hline
Spectroscopic Parameters: \\
$[$Fe/H] & Metallicity (dex) & 0.29 ± 0.08 & This work (HARPS-N) \\
$T_\mathrm{eff}$ & Effective Temperature (K) & 5493 ± 50 & This work (HARPS-N) \\
$\log(g)$ & Surface gravity (cgs) & 4.49 ± 0.10 & This work (HARPS-N) \\
$v\sin i_\star$ & Rotational velocity (km~s$^{-1}$) & <2  & This work (HARPS-N)  \\
\hline
\enddata
\tablecomments{References are: TIC \citep{stassun2018}, Tycho-2 \citep{Hog2000}, 2MASS  \citep{Cutri2003}, AllWISE \citep{Cutri2021}, Gaia DR3 \citep{GaiaCollaboration2022}. Gaia DR3 RA and Dec have been corrected from epoch J2016 to J2000. $v\sin i_\star$ has not been corrected for macroturbulence, and is therefore larger than the true $v\sin i_\star$. Floor errors have been adopted on $[\mathrm{Fe/H]}$, $T_\mathrm{eff}$, and $\log(g)$ to account for residual systematic errors.}
\end{deluxetable*}

One important feature of this intermediate-mass population is its compositional diversity, which (at least in a bulk sense) can be inferred when both masses and radii are well-measured \citep{Lopez2014, Thorngren2016}. Transiting planets with $20M_\Earth < M_\mathrm{p} < 100 M_\Earth$ span a wide range of sizes, indicating a wide range of compositions for the population \citep{Petigura2017}. Their compositions are broadly consistent with expectations from formation models except for a striking population of low-mass low-density outliers, sometimes called ``super-puffs'' \citep{Lee2016, Lee2019}. These mysteriously low-density ($\rho \lesssim 0.2$~g~cm$^{-3}$) planets were unanticipated by formation models, as they appear to have accreted voluminous H/He envelopes despite having smaller cores than typically required for runaway gas accretion. 

Such low-density outcomes of planet formation are still not fully understood, in part because they are rare. There are only 15 planets in this intermediate mass regime ($20M_\Earth < M_\mathrm{p} < 100 M_\Earth$) with densities below $\rho \leq 0.2$~g~cm$^{-3}$ \citep[per the NASA Exoplanet Archive on 23 March 2023;][]{Akeson2013}, and many reside in systems too faint for precise characterization. Detecting new low-density worlds in bright systems will help enable comparative planetology in this puzzling population. The Transiting Exoplanet Survey Satellite \citep[TESS;][]{Ricker2015} is playing an important role in detecting new puffy planets in systems amenable to follow-up efforts \citep[e.g.][]{McKee2022}.

To this end, we report the discovery of an exceptionally low-density ($\rho = 0.08 \pm0.02$~g~cm$^{-3}$) planet orbiting the late G dwarf star TOI-1420 every 6.96 days. The planet TOI-1420b has a size similar to that of Jupiter ($R_\mathrm{p} = 11.9\pm0.3R_\Earth$) but a mass similar to that of Neptune ($M_\mathrm{p} = 25.1\pm3.8M_\Earth$). In Section~\ref{sec:obs} we describe the TESS observations that revealed the initial transit signals, as well as the follow-up photometric, spectroscopic, and imaging observations that ultimately confirmed the planet. In Section~\ref{sec:jointfit}, we present a global fit to the aforementioned observations using \texttt{EXOFASTv2}. We then examine the structure of this intriguingly low-density planet in Section~\ref{sec:disc}, and finally we conclude with a look towards future observations in Section~\ref{sec:conc}.

\section{Observations} \label{sec:obs}

\subsection{TESS Photometry} \label{sec:tess}

TOI-1420 (stellar parameters provided in Table~\ref{tab:stellar}) was selected for 2-minute cadence observations with the TESS mission starting in Sector 16. Raw photometric data were processed by the TESS Science Processing Operations Center \citep[SPOC;][]{jenkins2016}, based at the NASA Ames Research Center, and the resulting light curves were available to download from the Mikulski Archive for Space Telescopes (MAST). The SPOC conducted a transit search of Sector 16 on 2019 October 22 with an adaptive, noise-compensating matched filter \citep{Jenkins2002, Jenkins2010}, producing a threshold-crossing event (TCE) for which an initial limb-darkened transit model was fitted \citep{Li2019} and a suite of diagnostic tests were conducted to help make or break the planetary nature of the signal \citep{Twicken2018}. The TESS Science Office (TSO) reviewed the vetting information and issued an alert on 2019 November 6 \citep{Guerrero2021}. The signal was repeatedly recovered as additional observations were made in sectors 24, 56, 57 and 58. The transit signature passed all the diagnostic tests presented in the Data Validation reports. According to the difference image centroiding tests, the host star is located within $0\farcs339\pm2\farcs61$ of the transit signal source.

\begin{figure*}[t]
\begin{center}
% \epsscale{1}
\includegraphics[width=\textwidth]{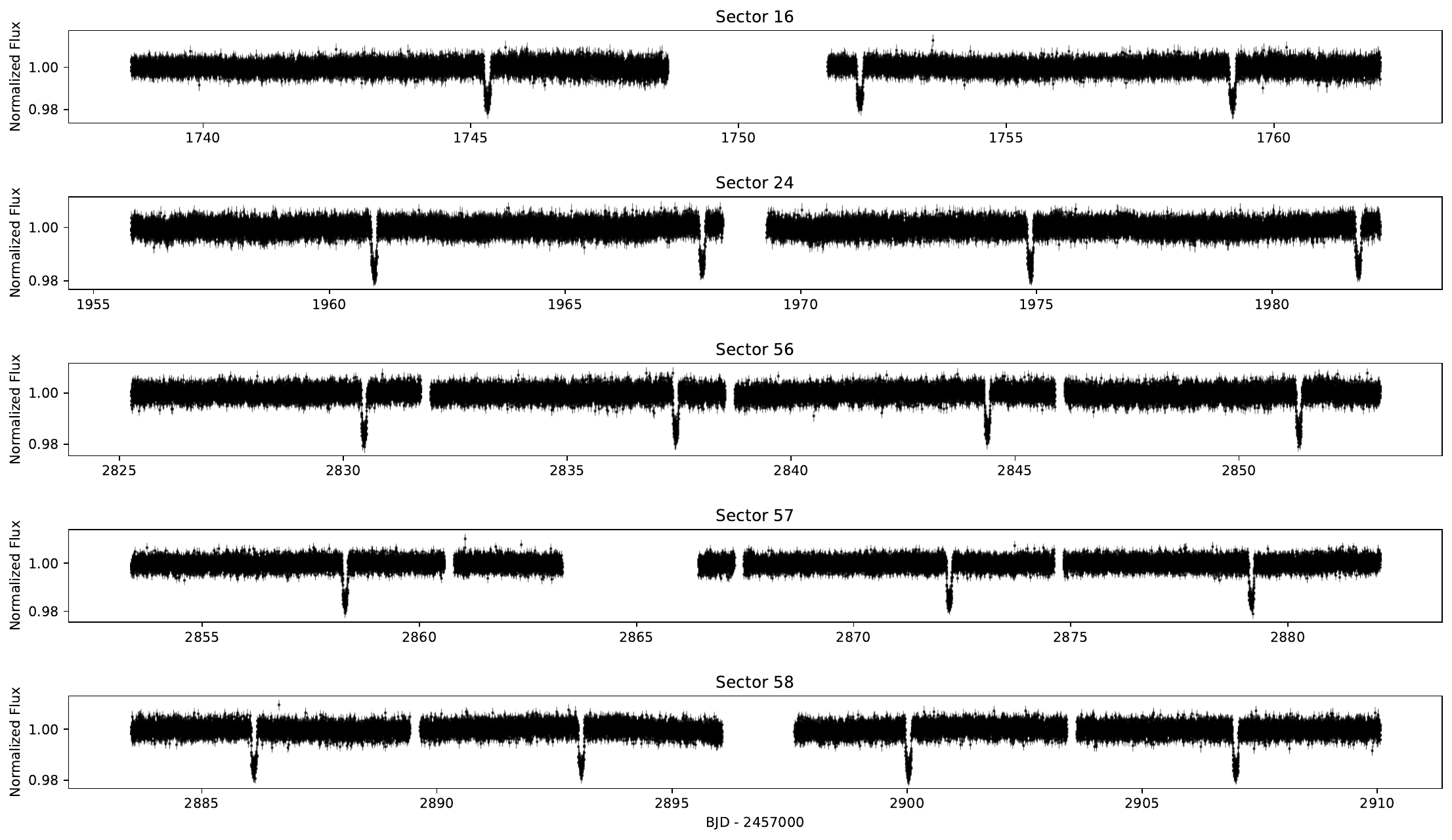}
\caption{TESS 2~min cadence PDCSAP light curves for TOI-1420. Transit events are visible roughly every 7~days.}
\label{fig:raw_tess}
% \end{minipage}
\end{center}
\end{figure*}

In our analysis, we included the 2~minute cadence data from Sectors 16, 24, 56, 57, and 58 (Figure~\ref{fig:raw_tess}). Data from Sectors 17 and 18 were taken only at 30~minute cadence and were not included. For the five 2~minute cadence Sectors, we obtained the Presearch Data Conditioning Simple Aperture Photometry (PDCSAP) light curves \citep{Smith2012, Stumpe2012, Stumpe2014} using the \texttt{lightkurve} package \citep{lightkurve}. We used \texttt{lightkurve} to detrend the light curves with a Savitsky-Golay filter, applying a window length of 66.7 hours (with the transit events masked, each with a duration of 3.37 hours), and we subsequently removed any $\geq5\sigma$ outliers from the light curves.

\begin{figure*}[t]
\begin{center}
% \epsscale{1}
\includegraphics[width=\textwidth]{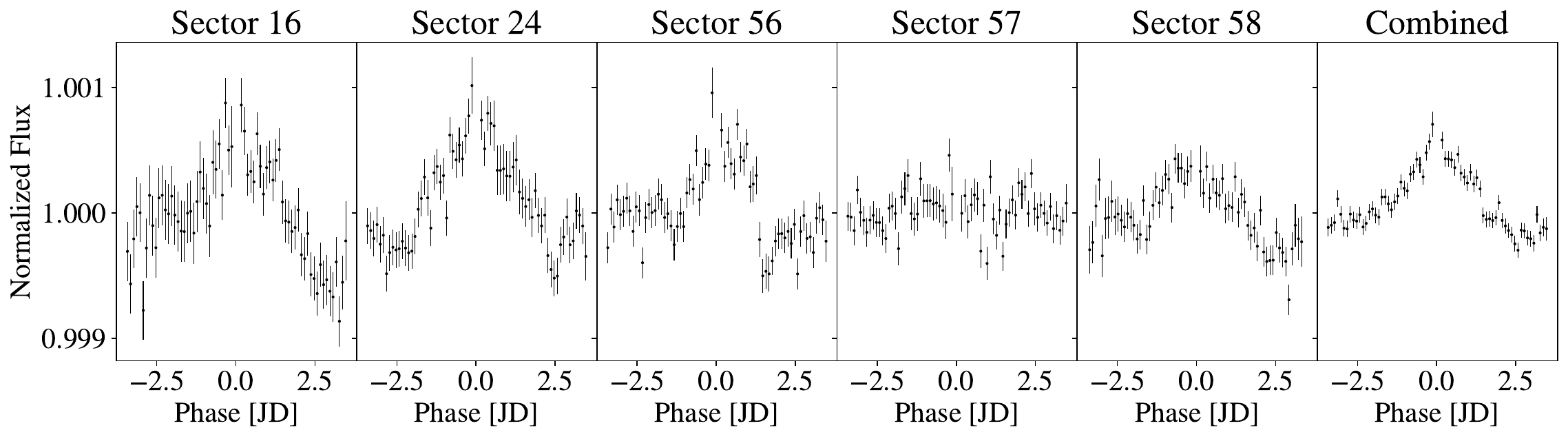}
\caption{TESS PDCSAP light curves for TOI-1420 phased to the planetary ephemeris and binned to 0.1~day cadence. The transits are masked and occur at a phase of 0~days.}
\label{fig:rotation}
% \end{minipage}
\end{center}
\end{figure*}

While detrending the TESS light curves, we noticed that some of the sectors exhibit periodic photometric variability on a 7~day timescale (close to the planetary orbital period); this was confirmed via a Lomb-Scargle periodogram and autocorrelation analysis. Moreover, in these cases the planetary transits seem to be phased up with the $\sim$500~ppm photometric variability. We show in Figure~\ref{fig:rotation} that the periodic variability appears fairly persistent across most TESS sectors, although the signal is not detected in Sector 57 and is fairly weak in Sector 58. Given this, it was especially important to use ground-based follow-up observations to test whether an on-target eclipsing binary (EB) or nearby eclipsing binaries (NEBs) blended with the target star within the TESS aperture were responsible for the observed transit signals in TESS. We note that the system is currently included in the TESS Eclipsing Binary catalog albeit with an ``ambiguous'' disposition \citep{Prsa2022}. 

\subsection{Ground-based Photometry} \label{groundbased}

 \begin{deluxetable}{ccccccc}
 \tablecaption{Summary of Ground-Based Light Curve Observations \label{tab:followup}}
 \tablecolumns{6}
 \tablehead{
 \colhead{Date} &
 \colhead{Observatory} &
 \colhead{Filter} &
 \colhead{Coverage} &
 \colhead{Size} &
  \colhead{Pixel Scale} \\
  \colhead{(UT)} & & & & \colhead{(m)} & \colhead{($\arcsec$)}
 }
 \startdata
 2019-11-27 & LCOGT & $I$ & Ingress & 2.0 & 0.30\\
 2020-02-04 & OPM & $B$ & Full & 0.2 & 0.69 \\
 2020-11-08 & Whitin & $z'$ & Full & 0.7 & 0.67 \\
 2021-01-03 & Wendelstein & $i$ & Full & 0.4 & 0.64 \\
 2021-05-15 & OAUV & $B$ & Egress & 0.5 & 0.54 \\
 2021-11-06 & Ca l'Ou & $B$ & Full & 0.4 & 1.11 \\
 2022-06-23 & KeplerCam & $i'$ & Ingress & 1.2 & 0.67\\
 2022-08-11 & LCOGT & $B$ & Full & 1.0 & 0.39 \\
 2022-08-11 & LCOGT & $g'$ & Full & 1.0  & 0.39 \\
 2022-08-11 & LCOGT & $i'$ & Full & 1.0  & 0.39 \\
 2022-08-11 & LCOGT & $z_s$ & Full & 1.0  & 0.39  \\
 \enddata
 \end{deluxetable}
 
The TESS pixel scale is $\sim 21\arcsec$ pixel$^{-1}$ and photometric apertures typically extend out to roughly 1 arcminute, generally causing multiple stars to blend in the TESS aperture. To rule out an NEB blend as the potential source of the TOI-1420.01 TESS detection and attempt to detect the signal on-target, we observed the field as part of the TESS Follow-up Observing Program\footnote{https://tess.mit.edu/followup} Sub Group 1 \citep[TFOP;][]{collins:2019}. We observed in multiple bands across the optical spectrum to check for wavelength dependent transit depth, which can also be suggestive of a planet candidate false positive. We used the {\tt TESS Transit Finder}, which is a customized version of the {\tt Tapir} software package \citep{Jensen:2013}, to schedule our transit observations. All light curve data are available on the {\tt EXOFOP-TESS} website\footnote{\href{https://exofop.ipac.caltech.edu/tess/target.php?id=321857016}{https://exofop.ipac.caltech.edu/tess/target.php?id=321857016}}.

In total, we obtained 11 follow-up observations from seven unique observatories: Observatoire Priv\'{e} du Mont (OPM), Whitin Observatory, Wendelstein Observatory, Observatori Astron\`{o}mic de la Universitat de Val\`{e}ncia (OAUV), Observatori de Ca l'Ou, KeplerCam, and the Las Cumbres Observatory Global Telescope \citep[LCOGT;][]{Brown:2013} 1~m and 2~m networks. Parameters for these follow-up observations are provided in Table~\ref{tab:followup}. All data were both calibrated and processed using {\tt AstroImageJ} \citep{Collins:2017} except for the LCOGT light curves, which were initially calibrated using the standard LCOGT {\tt BANZAI} pipeline \citep{McCully:2018, banzai}. 

\begin{figure}[h!]
\begin{center}
% \epsscale{1}
\includegraphics[width=0.48\textwidth]{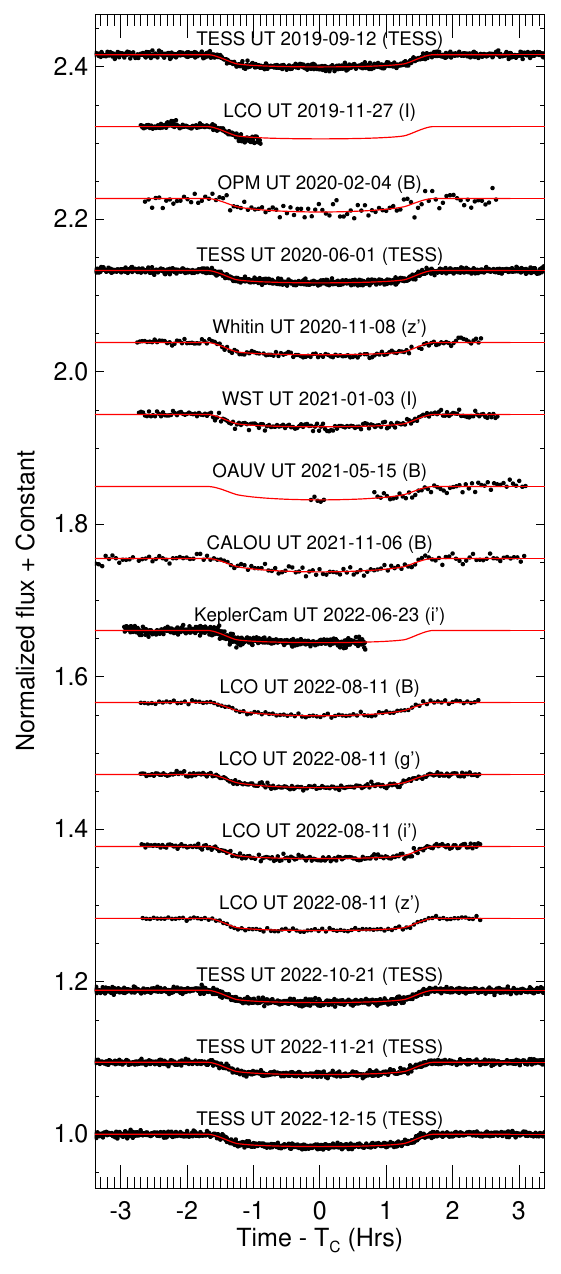}
\caption{Transit light curves of TOI-1420b with TESS and ground-based observatories. Red curves indicate the best-fit \texttt{EXOFASTv2} model.}
\label{fig:raw}
% \end{minipage}
\end{center}
\end{figure}

The transit light curves are shown in Figure~\ref{fig:raw}. In all cases, transit events were detected on-target, and on-time relative to the ephemerides from TESS. The depth of the detected events matched the depth in the TESS light curves, and the transits were achromatic (as determined by independent fits to the light curves, where the transit depths were all consistent with that observed by TESS). Thus, NEB blends were confidently ruled out as the source of the transit signal.

\subsection{Radial Velocity Observations} \label{rvs}
To measure the mass of TOI-1420.01 and/or rule out an on-target EB as the source of the transit events, we scheduled reconnaissance spectroscopy observations with the Tillinghast Reflector Echelle Spectrograph (TRES). TRES is a fiber-fed optical echelle spectrograph on the 1.5-meter Tillinghast telescope at the Fred Lawrence Whipple Observatory on Mt. Hopkins in Arizona with a resolving power of 44,000 \citep{Szentgyorgyi2007}. Eight measurements were taken between 2019 December 6 and 2021 June 27 with exposure times ranging from 1200~s to 2700~s. Though the planet was not massive enough to be detected by TRES with radial velocities (RVs), the non-detection ruled out the possibility of an EB. 

The TRES observations indicated the spectrum of TOI-1420b was well suited for precise radial-velocity observations using the High Accuracy Radial velocity Planet Searcher for the Northern hemisphere (HARPS-N) at the 3.6m Telescopio Nazionale Galileo (TNG) in La Palma, Spain \citep{cosentino2012, Cosentino2014}. HARPS-N is a highly stabilized echelle spectrograph with a resolving power of $R\sim115,000$ capable of measuring radial velocities in the m~s$^{-1}$ regime. We observed TOI-1420 between 2021 October 25 and 2022 September 5 (Table~\ref{tab:rvs}) and amassed a total of 44 observations using 1800~s exposures. We extracted RVs from these observations using v2.3.5 of the HARPS-N Data Reduction Software, which cross-correlates each observed spectrum with a weighted binary mask to estimate the RV \citep{Pepe2002, Dumusque2018}. Five observations were removed either due to their low SNRs ($<20$) or due to the Rossiter-McLaughlin effect during transits \citep{ohta2005}. The final data set had internal precisions ranging from 1.9 to 5.4 m~s$^{-1}$, in line with the anticipated precisions. A Lomb-Scargle periodogram of the HARPS-N RVs alone revealed a signal at 6.9~days, consistent with the orbital period obtained from the TESS light curve. We also verified that the phasing of the RVs was independently consistent with the reported transit ephemeris. We used the HARPS-N spectra to constrain the stellar effective temperature $T_\mathrm{eff}$, surface gravity $\log(g)$, metallicity [Fe/H], and projected rotational velocity $v\sin i$ using the Stellar Parameter Classification code \citep[SPC;][]{Buchhave2012, Bieryla2021}. These measurements are reported in Table~\ref{tab:stellar} along with other stellar parameters from the literature.

We also obtained 14 RV measurements with the NN-explore Exoplanet Investigations with Doppler spectroscopy (NEID) instrument, a high-resolution ($R\sim110,000$) spectrograph at the WIYN 3.5-meter telescope\footnote{WIYN is a joint facility of the University of Wisconsin–Madison, Indiana University, NSF’s NOIRLab, the Pennsylvania State University, and Purdue University.} on Kitt Peak, Arizona \citep{Halverson2016, Schwab2016, Stefansson2016, Kanodia2018, Robertson2019}. We obtained 990~s exposures with the high resolution fiber between 2022 April 2 and 2022 June 6. The standard NEID data reduction pipeline\footnote{https://neid.ipac.caltech.edu/docs/NEID-DRP/} was used to obtain RVs from these observations, which are included in Table~\ref{tab:rvs}. RV precisions ranged from 2.5 to 7.5~m~s$^{-1}$, in line with the anticipated precision of 3.2~m~s$^{-1}$ from the NEID Exposure Time Calculator and sufficient to detect the planetary signal at $>3\sigma$ confidence in the NEID data. We also re-derived the RVs using the SERVAL pipeline optimized for NEID \citep{Zechmeister2018, Stefansson2022}, and confirmed that the RV signal observed with NEID was robust to different data processing schemes.

\begin{figure*}[b]
    \centering
    \includegraphics[width=0.48\textwidth]{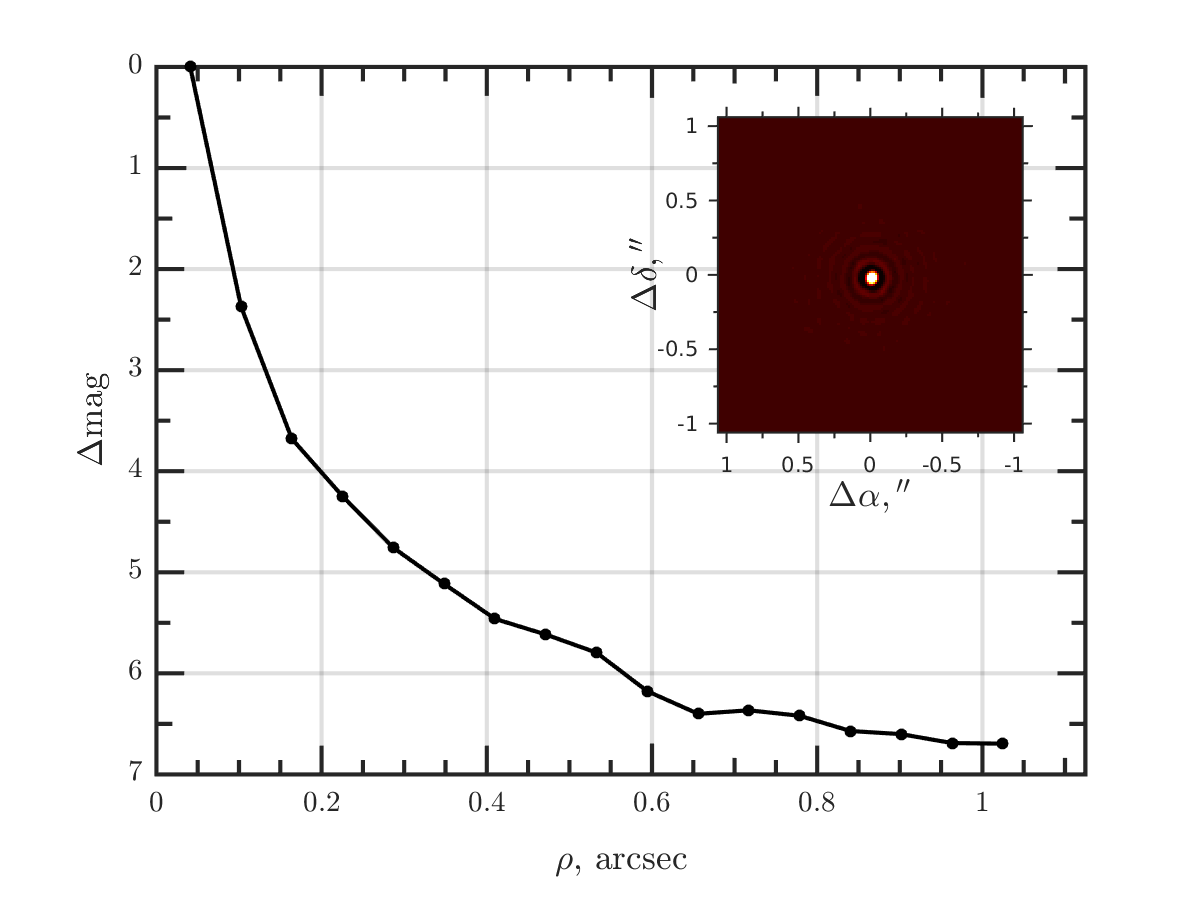}
    \includegraphics[width=0.48\textwidth]{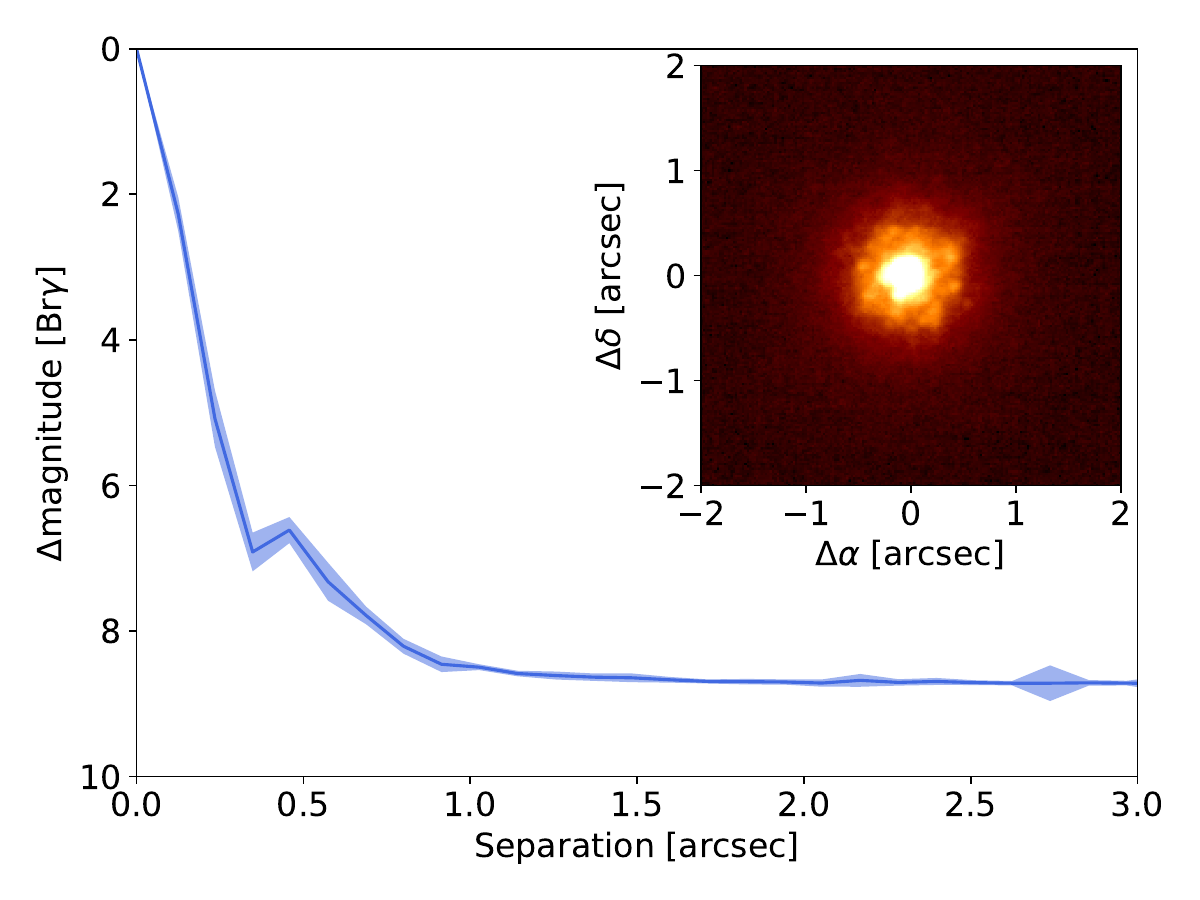}
    \caption{Sensitivities of our SAI-2.5~m $I_c$-band speckle image (left) and Gemini/NIRI Br$\gamma$ AO image (right). The insets are the high-resolution images of TOI-1420. The star appears single, and no visual companions are observed anywhere in either field of view, which extend to $5\farcs1\times10\farcs6$ for the SAI image and $26\arcsec\times26\arcsec$ for the Gemini/NIRI image (we only show the central few $\arcsec$ in the contrast curves and inset images for visual clarity).}
    \label{fig:highres}
\end{figure*}

\begin{figure*}[h]
% \begin{minipage}[b]{0.5\linewidth}
\centering
\includegraphics[width=0.48\textwidth]{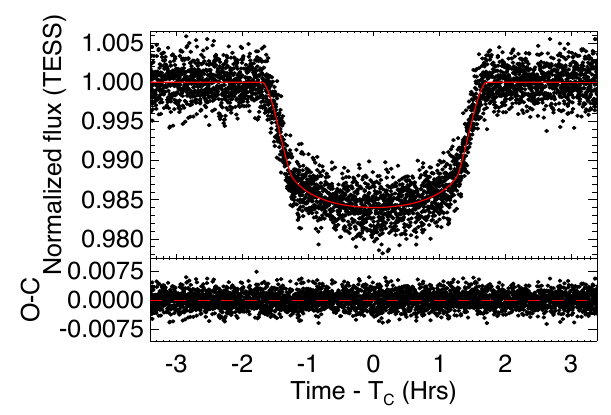}
\includegraphics[width=0.48\textwidth]{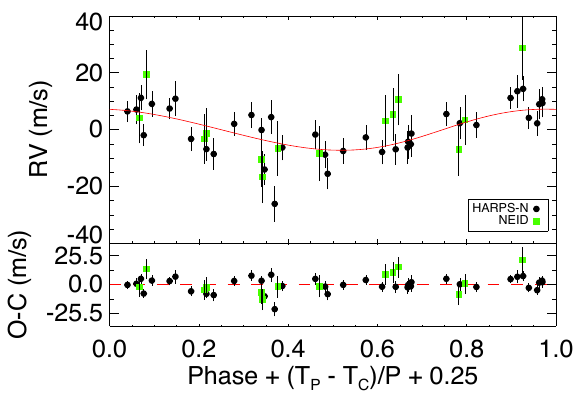}
\caption{Left: Phase-folded light curve of TESS transits (left) and phase-folded HARPS-N and NEID RVs of TOI-1420 (right) with residuals. The black data points are the observations and the red lines are the overlaying best-fit models from \texttt{ExoFASTv2}.}
\label{fig:fig6}
% \end{minipage}
\end{figure*}

\begin{deluxetable}{ccc}
\tablecaption{TOI-1420 RV Measurements from HARPS-N and NEID \label{tab:rvs}}
\tablecolumns{3}
\tablehead{
\colhead{BJD$_\mathrm{TDB}$} &
\colhead{RV (m~s$^{-1}$)} &
\colhead{$\sigma_\mathrm{RV}$ (m~s$^{-1}$)}
}
\startdata
HARPS-N Measurements: \\
2459513.43049 & -10332.6 & 3.2 \\
2459514.45289 & -10325.9 & 3.1 \\
2459515.39936 & -10325.3 & 2.8 \\
... & ... & ... \\
2459826.46509 & -10328.9 & 5.8 \\
2459828.50094 & -10316.8 & 2.8 \\
\hline
NEID Measurements: \\
2459736.87178 & -10253.7 & 2.8 \\
2459733.95347 & -10263.7 & 4.6 \\
2459732.84176 & -10258.4 & 3.5 \\
... & ... & ... \\
2459679.98135 & -10254.0 & 3.6 \\
2459672.00255 & -10265.5 & 4.9 \\
\enddata
\tablecomments{Table~\ref{tab:rvs} is published in its entirety in machine-readable format. A portion is shown here for guidance regarding its form and content.}
\end{deluxetable}

\subsection{High-Resolution Imaging} \label{imaging}

It is important to vet for close visual companions that can dilute the lightcurve and thus alter the measured radius, or cause false positives if the companion is itself an eclipsing binary \citep[e.g.][]{ciardi2015}. To search for nearby companions that are unresolved in TESS or in ground-based seeing-limited images, we obtained high-resolution images of TOI-1420. 

We observed TOI-1420 on 2020 December 2 UT with the Speckle Polarimeter \citep{Safonov2017} on the 2.5~m telescope at the Caucasian Observatory of Sternberg Astronomical Institute (SAI) of Lomonosov Moscow State University. The speckle polarimeter uses an Electron Multiplying CCD Andor iXon 897 as a detector. The atmospheric dispersion compensator allowed observation of this relatively faint target through the wide-band $I_c$ filter. The power spectrum was estimated from 4000 frames with 30 ms exposure. The detector has a pixel scale of $20.6$ mas~pixel$^{-1}$, and the angular resolution was 89~mas. We did not detect any stellar companions brighter than $\Delta I_C=4.5$ and $6.6$ at $\rho=0\farcs25$ and $1\farcs0$, respectively, where $\rho$ is the separation between the source and the potential companion. The speckle image of the target is shown in Figure~\ref{fig:highres} along with the 5$\sigma$ contrast curve.

We also vetted for close companions with adaptive optics (AO) imaging using Gemini/NIRI \citep{Hodapp2003}. We collected 9 science images on 2019 November 13, each with exposure time 18~s, using the Br$\gamma$ filter. The telescope was dithered between exposures in a grid pattern. We used the dither frames themselves to reconstruct a sky background, which was subtracted from all frames. We also corrected for bad pixels, flat-fielded, and then aligned frames based on the stellar position and co-added the stack of images. We finally determined the sensitivity of these observations as a function of radius by injecting fake companions, and scaling their brightness such that they are detected at 5$\sigma$. This was repeated at several radii and position angles, and sensitivities were averaged azimuthally. We do not detect companions anywhere within the field of view ($26\arcsec\times26\arcsec$ centered on TOI-1420). The data are sensitive to companions 5 magnitudes fainter than the star (=1\% flux dilution) beyond 232~mas from TOI-1420, and to companions 8.7 magnitudes fainter than the star in the background-limited regime. In Figure~\ref{fig:highres} we show the AO image of the target, as well as the sensitivity as a function of radius of these observations.

\subsection{Summary of Follow-Up Observations} \label{sec:confirmed}

With our follow-up data, we can confidently rule out false positive scenarios for TOI-1420.01. The seeing-limited photometry from Section~\ref{groundbased} localizes the transit signal to TOI-1420, which is not in a visual binary, and thus NEB blend scenarios are ruled out. The high-resolution imaging from Section~\ref{imaging} rules out a closer AO binary with multiple independent observations. Taken together with the low Gaia DR3 Renormalized Unit Weight Error (RUWE) of 0.866 (indicative of an acceptable single-star astrometric fit), these data indicate that TOI-1420 is indeed a single star. If the system were an EB, our RVs in Section~\ref{rvs} would have easily detected a stellar-mass object in a 7~day orbit around TOI-1420. Instead, they show that the object is only $25.1 \pm 3.8 M_\Earth$, as described in the following section. Thus we conclude that that object in orbit is unambiguously a planet. We hereafter refer to the planet as TOI-1420b and proceed to fit for the planetary parameters.

\begin{figure}
    \centering
    \includegraphics[width=0.44\textwidth]{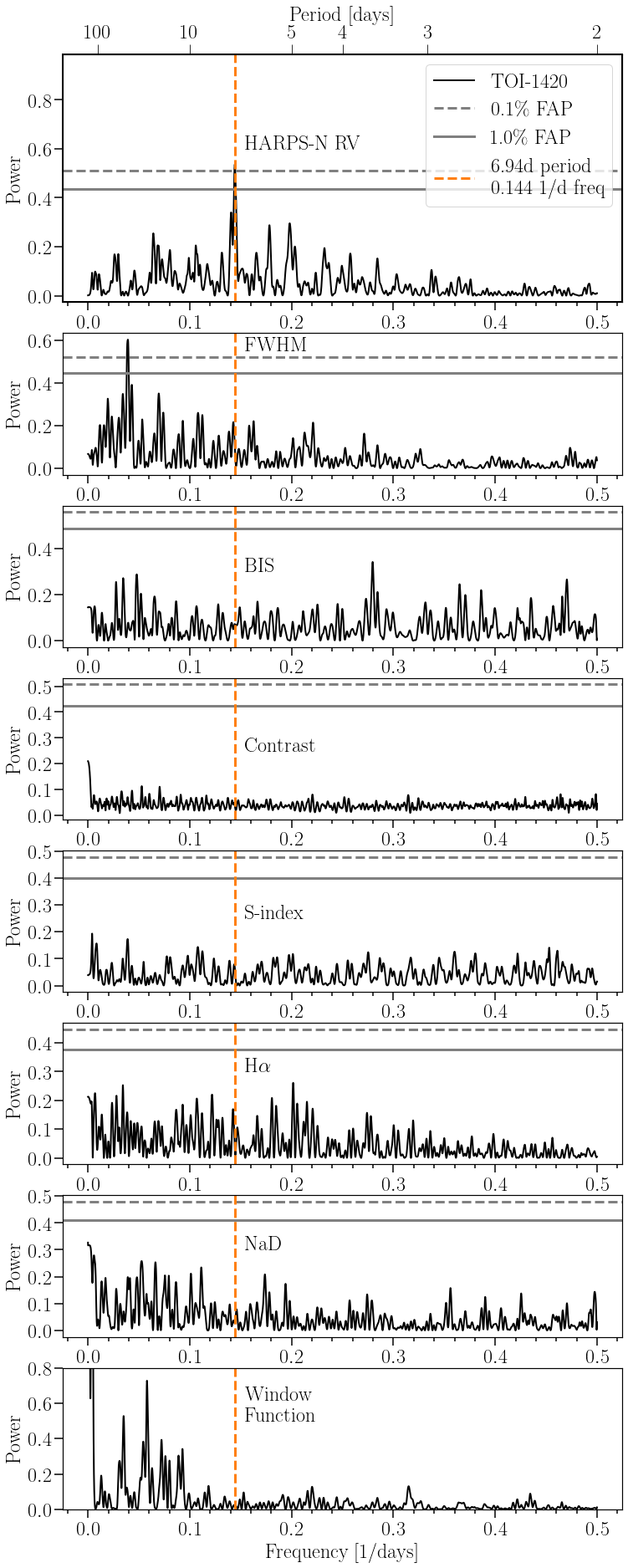}
    \caption{Periodograms of TOI-1420’s HARPS-N radial velocities, activity indicators, and window function. The False Alarm Probabilities (FAPs) are calculated via 100,000 bootstrap simulations. The only significant peak (<0.1\% FAP) in the RV periodogram is at 6.94 days, which we attribute to the planet candidate. This periodic signal is not present in any of the activity indicators.}
    \label{pgram}
\end{figure}

While we have ruled out EB false positive scenarios, the nature of the variability signal in the PDCSAP light curves shown in Figure~\ref{fig:rotation} remains somewhat unclear. We searched for similar variability in the PDCSAP light curves of nearby stars, but did not detect it, suggesting that the signal may not be instrumental in origin. The variability is not detected in ASAS-SN photometry of TOI-1420 \citep{Shappee2014, Kochanek2017}, but the photometric errors are too large (1-2\%) to be definitive. Next, we addressed the possibility that the variability was tracing stellar rotation. In Figure~\ref{pgram}, we show periodograms for a number of spectroscopic activity tracers from HARPS-N alongside a periodogram of the HARPS-N RVs. None of the activity tracers exhibit variations on a 7~d timescale, suggesting that this is not the true stellar rotation period.

To test whether this signal could be attributed to an uncorrected TESS systematic in the PDCSAP reduction, we used the Systematics-Insensitive Periodogram \citep[SIP;][]{Hedges2020}, which fits a linear noise model to the SAP light curves (with the transits masked) alongside a periodogram. We found that the SIP was unable to recover the 7~day periodicity, suggesting that the noise model was able to remove the variability signal. We conclude that the variability in Figure~\ref{fig:rotation} may be a TESS systematic that was left uncorrected by the PDCSAP pipeline. The orbital period of the TESS spacecraft is close to 14~days, nearly twice the orbital period of the planet, which may explain why the signal appeared to be correlated with the planetary orbit. In any case, the detrending procedure we applied in Section~\ref{sec:tess} removed the out-of-transit variations, minimizing any impact on our final inferred radius in the global fit.

\startlongtable
\begin{deluxetable*}{cccc}
\tablecaption{TOI-1420 Stellar and Planetary Properties \label{tab:mathmode}}
\tablecolumns{3}
\tablehead{
\colhead{Parameters} &
\colhead{Description (Units)} &
\colhead{Posterior Values}
}
\startdata
Stellar Parameters: \\
$M_*$ & Stellar Mass ($M_\odot$) & 0.987 ± 0.048 \\
$R_*$ & Stellar Radius ($R_\odot$) & 0.923 ± 0.024 \\
$L_*$ & Stellar Luminosity ($L_\odot$) & 0.705 ± 0.059 \\
$\rho_*$ & Stellar Density (g~cm$^{-3}$) & 1.77 ± 0.16 \\
$\log(g)$ & Stellar surface gravity (cgs) & 4.502 ± 0.029 \\
$[$Fe/H] & Metallicity (dex) & 0.280 ± 0.074 \\
$\varpi$ & Parallax (mas) &  4.951 ± 0.030 \\
$T_\mathrm{eff}$ & Effective Temperature (K) &  5510 ± 110 \\
Age & Age (Gyr) & $<10.7$ \\
$d$ & Distance (pc) & 202 ± 1.2 \\ 
$A_v$ &  $V$-band Extinction (mag) & 0.22 ± 0.11 \\
\hline
Planetary Parameters: \\
$P$ & Orbital Period (days) & 6.9561063 ± 0.0000017 \\
$T_0$ & Transit Epoch (BJD$_\mathrm{TDB}$) & 2459517.43305 ± 0.00012 \\
$R_p$ & Planetary Radius ($R_\earth$) & 11.89 ± 0.33 \\
$M_p$ & Planetary Mass ($M_\earth$) & 25.1 ± 3.8 \\
$\rho_p$ & Density (g~cm$^{-3}$) & 0.082 ± 0.015 \\
$R_p/R_\star$ & Planet-to-star radius ratio & 0.11816 ± 0.00059 \\
$a/R_\star$ & Semi-major axis/Stellar radius & 16.53 ± 0.47 \\
$\delta$ & Transit Depth (\%) & 1.396 ± 0.014 \\
$i$ & Orbital Inclination ($^{\circ}$) & 88.58 ± 0.13 \\
$a$ & Semi-major axis (au) & 0.0710 ± 0.0012 \\
$e$ & Eccentricity & $<0.17$ \\
$\omega_\star$ & Argument of periastron ($^{\circ}$) & -165 ± 77 \\
$K$ & RV Semi-amplitude (m~s$^{-1}$) & 8.5 ± 1.3 \\
$T_{14}$ &  Transit duration (days) & 0.1405 ± 0.0061 \\
$b$ & Impact Parameter & 0.412 ± 0.036 \\
$F$ & Incident Flux (Gerg~s$^{-1}$~cm$^{-2}$) & 0.189 ± 0.014 \\
$T_\mathrm{eq}$ & Equilibrium Temperature (K) & 957 ± 17 \\
\enddata
\tablecomments{Priors are as described in \texttt{EXOFASTv2} described in \citet{Eastman2019} with the addition of a metallicity prior from HARPS-N (Table~\ref{tab:stellar}) and a parallax prior from Gaia DR3 (Table~\ref{tab:stellar}) corrected for the bias reported by \citet{Lindegren2021}. We did not impose additional priors on the spectroscopic parameters $T_\mathrm{eff}$ and $\log(g)$, as these can suffer from systematic errors \citep[e.g.][]{Eastman2019}, so we fit them in ExoFAST independently to ensure we are not biased by systematic errors in the spectroscopy. Equilibrium temperature assumes zero albedo as described in \citet{Eastman2019}. Upper limits are at 2$\sigma$.}
\end{deluxetable*}

\section{Global Fit} \label{sec:jointfit}
We used the software package \texttt{EXOFASTv2} \citep{Eastman2019} to derive the stellar and planetary masses and radii from a joint solution of all the available photometry and spectroscopy. We used the TESS 2-minute cadence light curves, all ground based light curve observations, and the HARPS-N and NEID RV measurements. For the RV datasets, we fit for separate zero-point offsets and jitter values. We fit for the planetary radius, planetary mass, orbital inclination, orbital eccentricity, argument of periastron, orbital period, transit epoch, stellar temperature, stellar mass, stellar radius, stellar metallicity, stellar limb darkening coefficients visual extinction, distance, and parallax. The Markov Chain Monte Carlo (MCMC) fit was run with parallel tempering using 8 temperatures and 152 chains. We saved 7200 steps after thinning by a factor of 40. A burn-in period removed as described in Section~23.2 of \citet{Eastman2019}. We ensured that the Gelman-Rubin statistics for all parameters were $<$ 1.01 to indicate that the fit had sufficiently full convergence \citep{gelmanrubin1992}.

The results of this global fit are presented in Table~\ref{tab:mathmode}. Posteriors on additional fitting parameters including quadratic limb darkening coefficients, RV offsets and jitters, and added photometric variances are provided in the Appendix. We show the best-fit model with all ground and space-based light curves in Figure~\ref{fig:raw}. The phased TESS photometry, HARPS-N RVs, and NEID RVs are shown in Figure~\ref{fig:fig6} along with our best-fit solution. We find that the planet has a remarkably low density of just $0.082 \pm 0.015$~g~cm$^{-3}$. We verified the results of this global analysis by fitting the TESS photometry, HARPS-N RVs, and NEID RVs using the \texttt{exoplanet} package \citep{exoplanet:joss}. We used broad uniform priors on all parameters except for the stellar mass and radius, where we used the distributions from Table~\ref{tab:mathmode}. The posterior probability distributions from the \texttt{exoplanet} fit agreed with those from the \texttt{EXOFASTv2} fit to better than 1$\sigma$.

%\begin{figure*}[ht]
% \begin{minipage}[b]{0.5\linewidth}
%\centering
%\includegraphics[width=0.48\textwidth]{Screen Shot 2022-11-07 at 12.34.06 PM.png}
% \includegraphics[width=0.1\textwidth]{white image.png}
%\includegraphics[width=0.48\textwidth]{Screen Shot 2022-11-07 at 12.34.16 PM.png}
%\caption{The normalized, detrended light curves from TESS sectors 16 (left) and 24 (right). Detrending was performed in Keplerspline through EXOFASTv2.}
%\label{fig:fig5}
% \end{minipage}
%\end{figure*}

\section{Discussion} \label{sec:disc}

\begin{figure}
    \centering
    \includegraphics[width=0.45\textwidth]{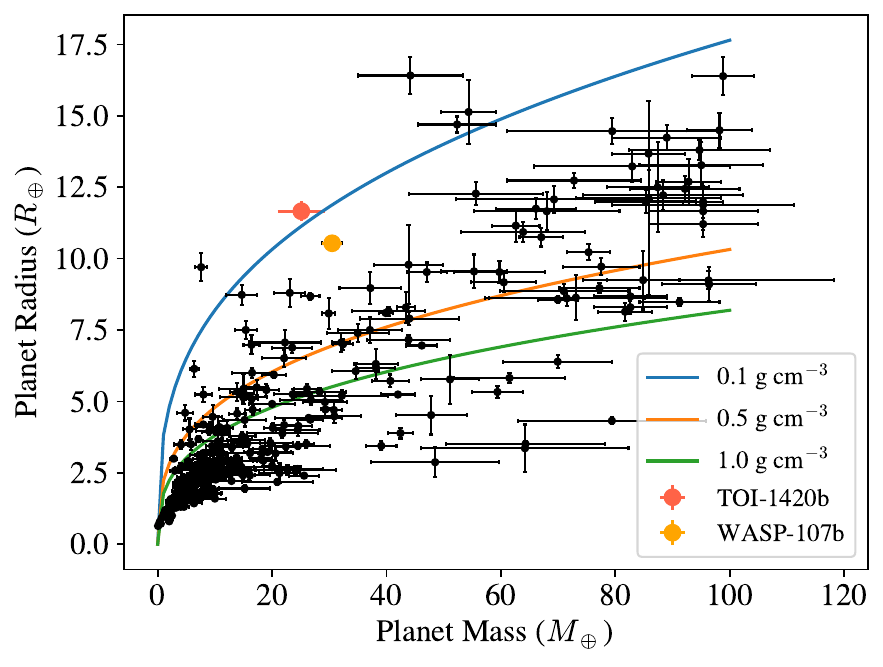}
    \caption{Mass-radius diagram of transiting planets from the NASA Exoplanet Archive with $0_\Earth < M_\mathrm{p} < 100M_\Earth$ and better than 30\% precision on the mass. The blue, orange, and green curves indicate 0.1~g~cm$^{-3}$, 0.5~g~cm$^{-3}$, and 1.0~g~cm$^{-3}$, respectively. TOI-1420 (red point) is extremely low density and falls along the 0.1~g~cm$^{-3}$ line.}
    \label{population}
\end{figure}

In Figure~\ref{population}, we present this new discovery in context on a mass-radius diagram of all known sub-Saturn mass planets (i.e. $20M_\Earth < M_p < 100 M_\Earth$) from the NASA Exoplanet Archive. The nearest neighbor to TOI-1420b on this plot is WASP-107b \citep{Anderson2017}, an important target for studies of planetary atmospheres, dynamics, structure, and formation \citep{Kreidberg2018, Spake2018, Piaulet2021, Rubenzahl2021}. TOI-1420b is larger and lower-mass than WASP-107b. Our newly-discovered planet also has a similar density to KELT-11b \citep{Pepper2017}, WASP-127b \citep{Lam2017}, and WASP-193b \citep{Barkaoui2023}, three puffy ($\rho\lesssim0.1$~g~cm$^{-3}$) planets with around twice the total mass of TOI-1420b. However, these three planets have equilibrium temperatures $T_\mathrm{eq} > 1000$~K and are thus likely to be inflated by the hot Jupiter radius inflation mechanism \citep[e.g.][]{Fortney2021}, whereas WASP-107b and TOI-1420b are too cool for substantial radius inflation.

Of these nearest neighbors, WASP-107b is particularly notable because it has an extreme envelope mass fraction of $>85\%$, corresponding to a low core mass of $<4.6M_\Earth$ \citep{Piaulet2021}. Given that TOI-1420b appears to be even more anomalous than WASP-107b, we constrained its bulk metallicity using a planetary structure model. We use the cool giant planet interior structure models of \citep{Thorngren2016} updated to use the \citet{Chabrier2019} equations of state for H and He. Matching these models to the observed parameters was done using the Bayesian framework described in \citet{Thorngren2019}. 

\begin{figure}
    \centering
    \includegraphics[width=0.5\textwidth]{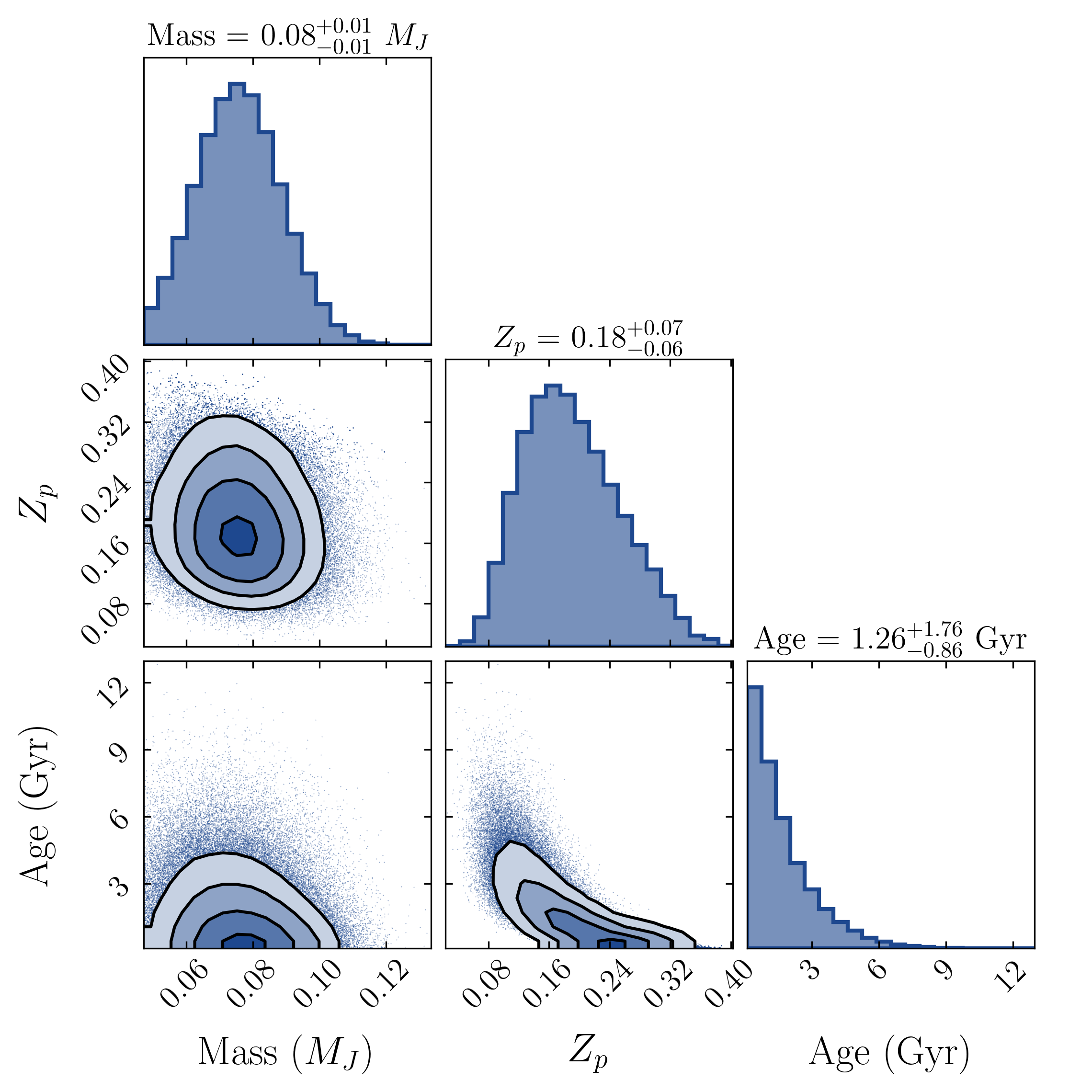}
    \caption{Posterior distributions on the planetary mass, bulk metallicity, and age from the planet evolution model described in the text.}
    \label{fig:metal}
\end{figure}

In Figure~\ref{fig:metal}, we show the posterior probability distribution for the bulk metallicity, where we find $Z_p = 0.18^{+0.07}_{-0.06}$. This corresponds to an envelope mass fraction $f_\mathrm{env} = 1 - Z_p = 0.82^{+0.07}_{-0.06}$ and an inferred core mass of at most $M_\mathrm{core} = M_pZ_p < 4.3^{+2.0}_{-1.7} M_\Earth$, similar to that of WASP-107b. The inferred core mass is an upper limit because the atmosphere may contain some metals (decreasing the amount available in the assumed core). There is a covariance with the age of the system (which is not well-constrained by our data): even with a larger metal fraction, a young planet would be puffy and match the observed radius. These calculations were run with no anomalous heating included, but because the planet is close to the hot Jupiter heating threshold of $F=2\times10^8$~erg~s$^{-1}$~cm$^{-2}$ \citep[see e.g.][]{Miller2011}, we re-ran the structure models with the radius decreased by 5\% to account for the possible weak heating. In this case we found $Z_p = 0.20\pm0.06$, i.e. weak anomalous heating does not appreciably change our inferred bulk metallicity.

We estimated the maximum atmospheric
metallicity (corresponding to an equally metal-rich atmosphere and interior) to be only $30\times$ Solar. The atmospheric metallicity was estimated by converting the 2$\sigma$ upper limit on bulk metallicity into a number fraction following Equation~(3) in \citet{Thorngren2019}, using a mean molecular mass for the metals of 18~amu (corresponding to water) and a helium to hydrogen mass ratio of 0.3383. This procedure assumes that the planet is fully mixed, which gives the largest (and therefore most conservative) upper limit on the atmospheric metallicity. The solar metallicity number ratio was taken to be $1.03\times10^{-3}$, multiplied by two to account for hydrogen being molecular in atmospheric conditions.

%However, the star does not appear to be unusually young. The star does exhibit a low-amplitude rotation period (see Section~\ref{sec:confirmed}), but it is not a member of any known moving group \citep[checked using BANYAN $\Sigma$;][]{Gagne2018}, the Li I 6708~\AA~line is not detected, and it has relatively low activity ($\log(R'_\mathrm{HK}) = -4.87$). These latter lines of evidence suggests that it is middle-aged. In this case, the core mass may be even smaller than what we have obtained here. 

Planets with such large envelopes despite their small cores are an interesting puzzle for core-nucleated accretion. In classic core-accretion models for planet formation, planets undergo runaway gas accretion when their cores grow to $\sim10M_\Earth$ \citep[e.g.][]{Pollack1996}. TOI-1420b and WASP-107b both appear to have accreted their envelopes with $\lesssim5M_\Earth$ cores. \citet{Stevenson1984} and \citet{Venturini2015} noted that runaway accretion can occur at relatively small core masses ($\lesssim2M_\Earth$) if the core forms water-rich beyond the ice line. On the other hand, \citet{Lee2016} suggest that such planets may form in ``dust-free'' regions of the disk, i.e. where the opacity is low and the planet can cool and accrete rapidly. Both of these scenarios require the planets to form further out before migrating inwards to their current positions. Alternatively, some investigators have proposed that the observed radii of low-density planets are inflated, either physically via tidal inflation \citep{Millholland2019, Millholland2020} or because high-altitude dust and/or hazes may set the photosphere at lower pressures than otherwise expected \citep{Kawashima2019, Wang2019, Gao2020}. Determining which (if any) of these mechanisms is at play for TOI-1420b will require more data: for instance, we could test the possibility of high-altitude dust/hazes \citep[e.g][]{Gao2020} and tidal heating \citep{Fortney2020} with a transmission spectrum, and additional RV observations could be used to search for a companion capable of driving the migration of TOI-1420b. Our current RV residuals are not yet sensitive to the presence of additional planets in the system. We also searched for transit-timing variations that could indicate another planet, but we found no detectable variations above a $2$~min amplitude.

\section{Conclusions and Future Work} \label{sec:conc}
We have confirmed TOI-1420b as an exceptionally low-density planet in a 6.96~day orbit around a late G dwarf. Using data from TESS, HARPS-N, NEID, and a number of other ground-based photometric and imaging facilities, we showed that the radius of this planet is 11.9 $\pm$ 0.3 $R_\Earth$ and the mass is 25.1 $\pm$ 3.8 $M_\Earth$. TOI-1420b is the largest known planet with a mass less than $50M_\Earth$, and it is similar to the planet WASP-107b in mass, radius, and irradiation. Using planetary structure models we showed that TOI-1420b has a large envelope mass fraction of 0.82$^{+0.07}_{-0.06}$, implying a core mass of only $M_\mathrm{core} \sim 4M_\Earth$. %The small core of TOI-1420b may have accreted its substantial envelope via dust-free accretion further from the star ($\gtrsim$1~au), with subsequent migration to its current position.

We encourage continued RV monitoring to further constrain the system architecture, which may reveal the dynamical history of TOI-1420b. An outer companion to WASP-107b was detected only after four years of Keck/HIRES monitoring \citep{Piaulet2021}, so similar long-term investments are warranted for the TOI-1420 system. Detecting an outer companion would strengthen the argument that close-in, low-density planets like WASP-107b and TOI-1420b form further out before dynamically migrating to their present positions \citep{Lee2016}. Additionally, Rossiter-McLaughlin (RM) constraints on the planet's sky-projected obliquity can also be informative about the migration history. Large obliquities \citep[like that observed for WASP-107b;][]{Dai2017, Rubenzahl2021} may be expected for planets that underwent scattering and/or high-eccentricity migration \citep[e.g][]{Dawson2018}. The misalignment can be damped via tidal interactions with the host star \citep{Albrecht2012}, but with $a/R_\star > 10$, TOI-1420b is unlikely to be significantly re-aligned by tides \citep{Rice2021}. Thus, if the system truly did have a dynamically-hot migration history, we may expect to observe a high obliquity. Using the upper limit on $v\sin i$ from Table~\ref{tab:stellar} and the transit parameters from Table~\ref{tab:mathmode}, the predicted RM amplitude for this system is \citep[e.g.][]{Triaud2018} $\frac{2}{3}(v\sin i_\star)\delta\sqrt{1 -b^2} < 17$~m~s$^{-1}$, well within reach for precise RV facilities assuming $v\sin i$ is close to our reported upper limit.

TOI-1420b is also an excellent target for atmospheric characterization. The planet's atmospheric scale height $H = \frac{kT}{\mu g}$ is 1950~km (assuming $\mu = 2.3$~amu), twice that of WASP-107b. TOI-1420b has a Transmission Spectroscopy Metric \citep[TSM; ][]{Kempton2018} of 580, where we have assumed the scale factor to be 1 as the scale factors in \citet{Kempton2018} are defined only up to 10$R_\Earth$. This puts the planet in rare company: TOI-1420b is the seventh-best exoplanet for transmission spectroscopy based on TSM, behind only WASP-107b, HD 209458b, HD 189733b, WASP-127b, KELT-11b, and WASP-69b. Low-density planets are also good targets for upper atmospheric characterization, as they are more susceptible to outflows than higher-gravity planets. We present evidence for helium in the upper atmosphere of TOI-1420b in a companion paper (Vissapragada et al. submitted). 

In all, TOI-1420b presents a number of exciting future prospects for atmospheric and dynamical characterization. Comparative planetology of TOI-1420b, WASP-107b, and other similarly low-density worlds will ultimately help unveil their formation and evolution histories.
\acknowledgements

We thank the referee for detailed comments that improved the quality of this manuscript, and we thank Andrew Vanderburg for helpful discussions regarding the TESS data. The postdoctoral fellowship of KB is funded by F.R.S.-FNRS grant T.0109.20 and by the Francqui Foundation. KAC acknowledges support from the TESS mission via subaward s3449 from MIT. KKM acknowledges support from the New York Community Trust Fund for Astrophysical Research. NH was supported by a NASA Massachusetts Space Grant fellowship. M.V.G. and I.A.S. acknowledge the support of Ministry of Science and Higher Education of the Russian Federation under the grant 075-15-2020-780 (N13.1902.21.0039). GS acknowledges support provided by NASA through the NASA Hubble Fellowship grant HST-HF2-51519.001-A awarded by the Space Telescope Science Institute, which is operated by the Association of Universities for Research in Astronomy, Inc., for NASA, under contract NAS5-26555.

We acknowledge the use of public TESS data from pipelines at the TESS Science Office and at the TESS Science Processing Operations Center. Resources supporting this work were provided by the NASA High-End Computing (HEC) Program through the NASA Advanced Supercomputing (NAS) Division at Ames Research Center for the production of the SPOC data products. This paper includes data collected by the TESS mission, which are publicly available from the Mikulski Archive for Space Telescopes (MAST) operated by the Space Telescope Science Institute (STScI). Funding for the TESS mission is provided by NASA’s Science Mission Directorate. The specific observations analyzed can be accessed via\dataset[10.17909/har4-7y03]{https://doi.org/10.17909/har4-7y03}.

This work makes use of observations from the LCOGT network. Part of the LCOGT telescope time was granted by NOIRLab through the Mid-Scale Innovations Program (MSIP). MSIP is funded by NSF. This research has made use of the Exoplanet Follow-up Observation Program (ExoFOP; DOI: 10.26134/ExoFOP5) website, which is operated by the California Institute of Technology, under contract with the National Aeronautics and Space Administration under the Exoplanet Exploration Program. Funding for the TESS mission is provided by NASA's Science Mission Directorate. This paper contains data taken with the NEID instrument, which was funded by the NASA-NSF Exoplanet Observational Research (NN-EXPLORE) partnership and built by Pennsylvania State University. NEID is installed on the WIYN telescope, which is operated by the National Optical Astronomy Observatory, and the NEID archive is operated by the NASA Exoplanet Science Institute at the California Institute of Technology. NN-EXPLORE is managed by the Jet Propulsion Laboratory, California Institute of Technology under contract with the National Aeronautics and Space Administration. Data presented herein were obtained at the WIYN Observatory, or the CTIO SMARTS 1.5m, or MINERVA-Australis from telescope time allocated to NN-EXPLORE through the scientific partnership of the National Aeronautics and Space Administration, the National Science Foundation, and the NOIRLab. The authors are honored to be permitted to conduct astronomical research on Iolkam Du’ag (Kitt Peak), a mountain with particular significance to the Tohono O’odham. This work has made use of data from the European Space Agency (ESA) mission {\it Gaia} (\url{https://www.cosmos.esa.int/gaia}), processed by the {\it Gaia} Data Processing and Analysis Consortium (DPAC,
\url{https://www.cosmos.esa.int/web/gaia/dpac/consortium}). Funding for the DPAC has been provided by national institutions, in particular the institutions participating in the {\it Gaia} Multilateral Agreement. This research made use of \texttt{exoplanet} \citep{exoplanet:joss,
exoplanet:zenodo} and its dependencies \citep{exoplanet:agol20,
exoplanet:arviz, exoplanet:astropy13, exoplanet:astropy18, exoplanet:kipping13,
exoplanet:luger18, exoplanet:pymc3, exoplanet:theano}.

\facilities{Exoplanet Archive, ExoFOP, TESS, TNG (HARPS-N), WIYN (NEID), LCOGT, FLWO:1.2m (KeplerCam), WCWO:0.6m, Gemini:Gillett (NIRI)}
\software{AstroImageJ \citep{Collins:2017}, TAPIR \citep{Jensen:2013}, EXOFASTv2 \citep{Eastman2019}, exoplanet \citep{exoplanet}, pymc3 \citep{exoplanet:pymc3}, numpy \citep{numpy}, scipy \citep{scipy}, astropy \citep{exoplanet:astropy13, exoplanet:astropy18}, BANZAI \citep{McCully:2018, banzai}}

\appendix

In Table~\ref{tab:extras}, we give the posteriors on RV offsets, added photometric variances, RV jitters, and quadratic limb darkening coefficients from our \texttt{ExoFASTv2} fit.

%\startlongtable
\begin{deluxetable*}{cccc}
\tablecaption{Additional posterior values from \texttt{ExoFASTv2} fit. \label{tab:extras}}
\tablecolumns{3}
\tablehead{
\colhead{Parameters} &
\colhead{Description (Units)} &
\colhead{Posterior Values}
}
\startdata
$\gamma_0$ & HARPS-N RV offset (m~s$^{-1}$) & -10327.60 ± 0.96 \\
$\gamma_1$ & NEID RV offset (m~s$^{-1}$) & -10255.2 ± 2.9 \\
$\sigma_{\mathbf{J},0}$ & HARPS-N Jitter (m~s$^{-1}$) & 4.75 ± 0.96 \\
$\sigma_{\mathbf{J},1}$ & NEID Jitter (m~s$^{-1}$) & 9.7 ± 3.1 \\
$\sigma^{2}_{0}$ & TESS Added Variance &
-0.00000351 ± 0.00000011 \\
$\sigma^{2}_{1}$ & LCO $I$ Added Variance &
0.00000725 ± 0.00000089 \\
$\sigma^{2}_{2}$ & OPM Added Variance &
0.0000377 ± 0.0000061 \\
$\sigma^{2}_{3}$ & TESS Added Variance &
-0.000001935 ± 0.000000087 \\
$\sigma^{2}_{4}$ & Whitin Added Variance &
0.00000525 ± 0.00000080 \\
$\sigma^{2}_{5}$ & WST Added Variance &
0.00000703 ± 0.00000081 \\
$\sigma^{2}_{6}$ & OAUV Added Variance &
0.0000150 ± 0.0000040 \\
$\sigma^{2}_{7}$ & Calou Added Variance &
0.0000114 ± 0.0000016 \\
$\sigma^{2}_{8}$ & KeplerCam Added Variance &
-0.00009439 ± 0.00000061 \\
$\sigma^{2}_{9}$ & LCO $B$ Added Variance &
0.00000074 ± 0.00000035 \\
$\sigma^{2}_{10}$ & LCO $g'$ Added Variance &
0.00000154 ± 0.00000035 \\
$\sigma^{2}_{11}$ & LCO $i'$ Added Variance &
0.00000281 ± 0.00000051 \\
$\sigma^{2}_{12}$ & LCO $z_s$ Added Variance &
0.00000055 ± 0.00000029 \\
$\sigma^{2}_{13}$ & TESS Added Variance &
-0.000000222 ± 0.000000096 \\
$\sigma^{2}_{14}$ & TESS Added Variance &
-0.000000048 ± 0.000000098 \\
$\sigma^{2}_{15}$ & TESS Added Variance &
-0.000000090 ± 0.000000088 \\
$u_{1,0}$ & $B$ Quadratic Limb Darkening & 0.756 ± 0.036 \\
$u_{2,0}$ & $B$ Quadratic Limb Darkening & 0.070 ± 0.037 \\
$u_{1,1}$ & $I$ Quadratic Limb Darkening & 0.345 ± 0.036 \\
$u_{2,1}$ & $I$ Quadratic Limb Darkening & 0.266 ± 0.035 \\
$u_{1,2}$ & $g'$ Quadratic Limb Darkening & 0.671 ± 0.044 \\
$u_{2,2}$ & $g'$ Quadratic Limb Darkening & 0.122 ± 0.052 \\
$u_{1,3}$ & $i'$ Quadratic Limb Darkening & 0.345 ± 0.034 \\
$u_{2,3}$ & $i'$ Quadratic Limb Darkening & 0.235 ± 0.035 \\
$u_{1,4}$ & $z_s$ Quadratic Limb Darkening & 0.291 ± 0.033 \\
$u_{2,4}$ & $z_s$ Quadratic Limb Darkening & 0.257 ± 0.033 \\
$u_{1,5}$ & TESS Quadratic Limb Darkening & 0.356 ± 0.022 \\
$u_{2,5}$ & TESS Quadratic Limb Darkening & 0.243 ± 0.023 \\
\enddata
\end{deluxetable*}

\clearpage
\bibliography{references}

\end{document}